\documentstyle[epsf,bibnorm,float]{lamuphys}
\makeatletter
\let\chapter\hid@chapter
\makeatother
\begin{document}
\pagestyle{empty}

\title{Realistic Shell-Model Calculations for $^{208}$Pb Neighbors}

\author{Luigi Coraggio, Aldo Covello, and Angela 
Gargano
}

\institute{Dipartimento di Scienze Fisiche, Universit\`a
di Napoli Federico II, and Istituto Nazionale di Fisica Nucleare,
Complesso Universitario di Monte S. Angelo, Via Cintia, I-80126 Napoli, Italy}

\maketitle

\begin{abstract}
We have performed a shell-model study of the two nuclei $^{210}$Po and 
$^{206}$Hg, having and lacking two protons with respect to doubly magic 
$^{208}$Pb. 
In our calculations we have employed realistic effective interactions derived 
from the Bonn A nucleon-nucleon interaction.
The calculated results are compared with the available experimental data which 
are, however, very scanty for $^{206}$Hg.
The very good agreement obtained for $^{210}$Po supports confidence in our 
predictions for $^{206}$Hg.
\end{abstract}

\section{Introduction}
The region of nuclei around $^{208}$Pb has long been the subject of both 
experimental and theoretical studies.
Clearly, this is related to the fact that $^{208}$Pb is a very good doubly 
magic nucleus, so that the structure of neighboring nuclei, having or lacking 
nucleons with respect to it, can be appropriately described in terms of shell 
model.

In this work, we focus attention on the $N=126$ isotones $^{210}$Po and 
$^{206}$Hg, since nuclei with two valence particles or holes provide an ideal 
testing ground for the matrix elements of the two-body residual interaction.
In most of the calculations performed so far for these nuclei 
\cite{ma73,glaudemans85} empirical effective interactions have been used.
As early as some twenty-five years ago, however, a realistic effective 
interaction, derived from the Hamada-Johnston nucleon-nucleon ($NN$) 
\hbox{potential \cite{hj62}}, was employed in the works of Refs. 
\cite{kuo71,kuo72,kuo75} to calculate two-particle and two-hole states in the 
Pb region.
Since that time there has been substantial progress towards a microscopic 
approach to shell-model calculations starting from a free $NN$ potential.
This has concerned both the two basic ingredients involved in such an approach, 
namely the $NN$ potential and the many-body methods for deriving the 
model-space effective interaction.
These improvements have been incorporated into the present calculations, which 
are a part of an extensive study aimed at understanding the role of modern 
realistic interactions in the shell-model approach to the nuclear many-body 
problem \cite{andr96,andr97a,andr97b,coraggio98}.
More precisely, our effective interaction has been derived from the 
meson-theoretic Bonn A potential within the framework of a $G$-matrix 
folded-diagram method.

The paper is organized as follows. 
In Sec. II we give a brief description of our calculations. 
In Sec. III we present the results obtained for $^{210}$Po and $^{206}$Hg and 
compare them with experimental data.
In Sec. IV we draw some conclusions of our study.

\section{Outline of Calculations}
As already mentioned in the Introduction, we make use of a realistic effective 
interaction derived from the Bonn A free $NN$ potential. 
This was obtained using a $G$-matrix folded-diagram formalism, including 
renormalizations from both core polarization and folded diagrams. 
Since the valence-proton and \hbox{-neutron} orbits outside $^{208}$Pb are 
different, we have chosen the Pauli exclusion operator $Q_2$ in the $G$-matrix 
equation,

\begin{equation}
G(\omega)=V+VQ_2 \frac{1}{\omega-Q_2TQ_2}Q_2G(\omega)~~,
\end{equation}

\noindent
as specified by $(n_1,n_2,n_3)=(22,45,78)$ for the neutron orbits, and by 
$(n_1,n_2,n_3)=(16,36,78)$ for the proton orbits \cite{krenc76}. 
Here $V$ represents the $NN$ potential, $T$ denotes the two-nucleon kinetic 
energy, and $\omega$ is the so-called starting energy. 
We employ a matrix inversion method to calculate the above $G$ matrix in an 
essentially exact way \cite{krenc76,tsai72}.
The effective interaction $V_{\mathrm{eff}}$, which is energy independent, can 
be schematically written in the operator form as

\begin{equation}
V_{\mathrm{eff}} = \hat{Q} - \hat{Q'} \int \hat{Q} + \hat{Q'} \int \hat{Q} \int 
\hat{Q} - \hat{Q'} \int \hat{Q} \int \hat{Q} \int \hat{Q} + ~...~~,
\end{equation}

\noindent
where $\hat{Q}$ and $\hat{Q'}$ represent the $\hat{Q}$ box, composed of 
irreducible valence-linked diagrams, and the integral sign represents a 
generalized folding operation. 
We take the $\hat{Q}$ box to be composed of $G$-matrix diagrams through second 
order  in $G$; they are just the seven first- and second-order diagrams 
considered by Shurpin {\em et al.} \cite{kuo83}.
It should be mentioned that in $^{206}$Hg we treat protons as valence holes, 
which implies the derivation of a hole-hole effective interaction.
In the calculation of $V_{\mathrm{eff}}$ we use an isospin uncoupled 
represention, where protons and neutrons are treated separately. 
For the shell-model oscillator parameter we have used 6.88 MeV, as obtained 
from the expression $\hbar \omega=45A^{-1/3} - 25A^{-2/3}$ for $A=208$.
A detailed description of our derivation including more references can be found 
in Ref. \cite{andr97a}.

As regards the single particle energies, we have taken them from the 
experimental spectra of $^{209}$Bi and $^{207}$Tl \cite{martin91,martin93}.
Thus, for $^{210}$Po we have used the following values (in MeV): 
$\epsilon_{h_{9/2}}=0.0$, $\epsilon_{f_{7/2}}=0.896$, 
$\epsilon_{i_{13/2}}=1.609$, $\epsilon_{f_{5/2}}=2.826$, 
$\epsilon_{p_{3/2}}=3.119$, $\epsilon_{p_{1/2}}=3.633$, while for 
$^{206}$Hg the adopted single-hole spectrum is $\epsilon_{s_{1/2}}=0.0$, 
$\epsilon_{d_{3/2}}=0.351$, $\epsilon_{h_{11/2}}=1.348$, 
$\epsilon_{d_{5/2}}=1.683$, $\epsilon_{g_{7/2}}=3.474$.

\section{Results}
In Fig.1 we report all the experimental \cite{browne92,mann88} and calculated 
levels of $^{210}$Po up to about 3.2 MeV. 

\begin{figure}[H]
\centerline{\epsfbox{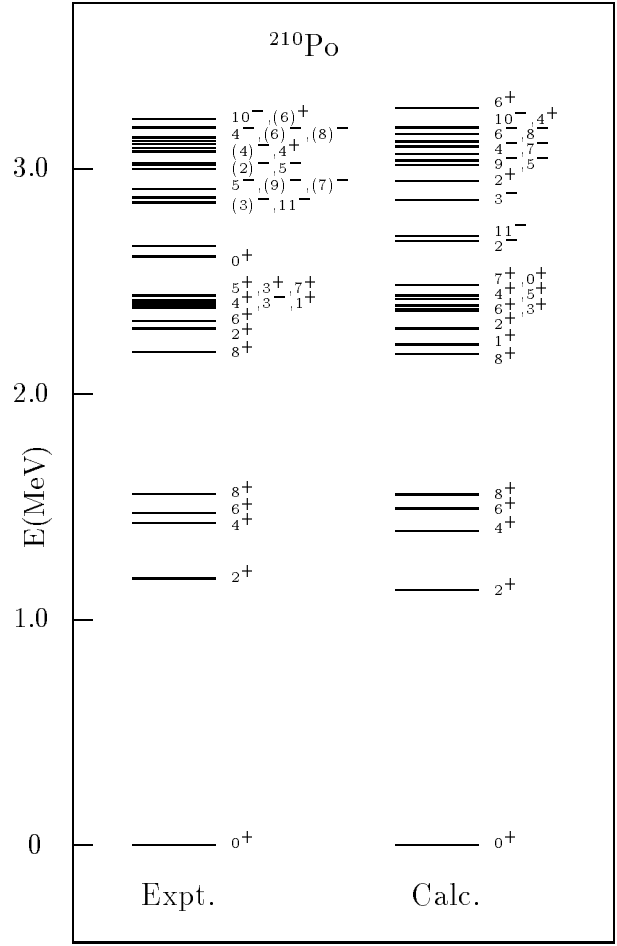}}
\caption[ ]{Experimental and calculated spectrum of $^{210}$Po.}
\label{210po}
\end{figure}

We see that each state of a given $J^{\pi}$ in the calculated spectrum has its 
experimental counterpart, the only exception being the $2^+_3$ state.
Experimentally, however, two levels with no angular momentum assignment 
have been observed at 2.658 and 2.872 MeV. 
One of these states may correspond to the theoretical $2^+_3$ at 2.947 MeV.
Three of the reported experimental levels, namely the $3^-_1$ state at 2.387 
MeV, the $5^-_1$ state at 2.910 MeV, and the $4^-_1$ state at 3.112 MeV, cannot 
be described within our model space; the first one reflects the collective 
nature of the octupole $3^-$ state at 2.615 MeV in $^{208}$Pb, while the other 
two levels arise from the neutron particle-hole configuration 
$\nu (g_{9/2}p^{-1}_{1/2})$ \cite{mann88}.
A measure of the quality of the results is given by the rms deviation $\sigma$ 
\cite{autoc}, whose value relative to the 25 identified excited states is 92 
KeV.

In Fig. 2 the observed \cite{helmer90} and theoretical spectra of $^{206}$Hg 
are reported.

\begin{figure}[H]
\centerline{\epsfbox{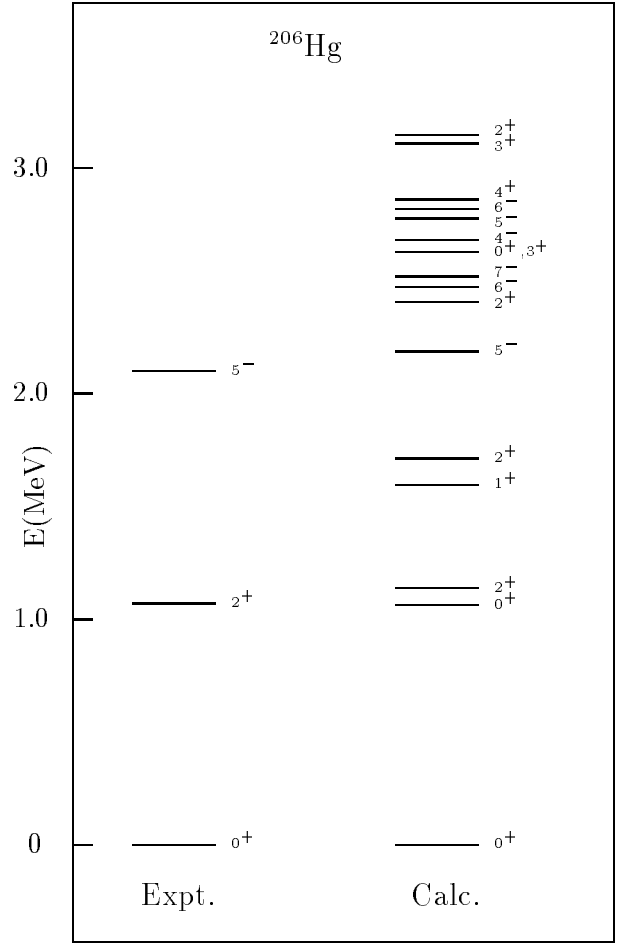}}
\caption[ ]{Experimental and calculated spectrum of $^{206}$Hg.}
\label{206hg}
\end{figure}

Only three excited states have been observed in this nucleus, and one of them, 
the $0^+_2$ at 3.625 MeV, is recognized to be a neutron pairing 
\hbox{vibration \cite{flynn78}}.
For this reason we have not reported this level in Fig.1.

We have also calculated the ground-state binding energies relative to 
$^{208}$Pb.
As for the Coulomb energy, we have taken that of a homogeneous charged 
sphere, whose radius is $R=r_0 A^{1/3}$, with $r_0=1.2$ fm.
We find $E_b$($^{210}$Po)=8.871 and $E_b$($^{206}$Hg)=$-15.165$ MeV, to be 
compared with the experimental values \cite{wapstra93} $8.783 \pm 0.004$ 
and $-15.382 \pm 0.021$ MeV, respectively.

\begin{table}[H]
\caption[]{Calculated and experimental reduced transition probabilities (in 
W.u.). The experimental data are from \cite{browne92,helmer90}.}
\begin{center}
\renewcommand{\arraystretch}{1.2}
\begin{tabular}{lccccccc}
\hline\noalign{\smallskip}
 Nucleus &~& $\lambda$ &~&$J^{\pi}_i \rightarrow J^{\pi}_f$ &~& 
\multicolumn{2}{c}{ $B(E\lambda$)} \\
 ~~~&~&~~~&~&~~~&~& Calc. & Expt. \\
\hline
 $^{210}$Po &~& 2 &~& $2_1^+ \rightarrow  0_1^+$ &~& 3.62  & $0.56 \pm 0.12$ \\
 ~~~~~~~~~  &~& 2 &~& $4_1^+ \rightarrow  2_1^+$ &~& 4.49  & $4.53 \pm 0.15$ \\
 ~~~~~~~~~  &~& 2 &~& $6_1^+ \rightarrow  4_1^+$ &~& 3.08  & $3.00 \pm 0.12$ \\
 ~~~~~~~~~  &~& 2 &~& $8_1^+ \rightarrow  6_1^+$ &~& 1.25  & $1.10 \pm 0.05$ \\
 ~~~~~~~~~  &~& 3 &~& $11_1^- \rightarrow 8_2^+$ &~& 7.5   & $19.7 \pm 1.1 $ \\
 ~~~~~~~~~  &~& 3 &~& $11_1^- \rightarrow 8_1^+$ &~& 0.53  & $3.71 \pm 0.10$ \\
 $^{206}$Hg &~& 2 &~& $2_1^+ \rightarrow  0_1^+$ &~& 5.2   & $> 0.00027$     \\
 ~~~~~~~~~  &~& 3 &~& $5_1^- \rightarrow  2_1^+$ &~& 0.432 & $0.182 \pm 0.018$ \\
\noalign{\smallskip}\hline
\end{tabular}
\renewcommand{\arraystretch}{1}
\label{table1}
\end{center}
\end{table}

In Table I the experimental reduced transitions probabilities in $^{210}$Po
and $^{206}$Hg \cite{browne92} are compared with the calculated ones.
We have used an effective proton charge $e^{\mathrm{eff}}_p=1.5~e$, which is 
consistent with the values adopted by other authors \cite{glaudemans85,kuo75}.
As regards $g_s$ and $g_l$, we have taken the values $g_s=3.5$ and $g_l=1.12$, 
which reproduce the $g$-factor of the $(\frac{9}{2}^-)_1$ state in $^{209}$Bi 
\cite{martin91} and the 
$B(M1;(\frac{3}{2}^+)_1 \rightarrow (\frac{1}{2}^+)_1$) in $^{207}$Tl 
\cite{martin93}.
The theoretical $B(E2)$ values are in good agreement with the observed ones, 
except for the $B(E2;2^+_1 \rightarrow 0^+_1$) in $^{210}$Po, which is 
overestimated by a factor of about six.
Our theoretical value, however, is consistent with that obtained by previous 
calculations \cite{glaudemans85}.
The calculated $B(E3)$'s in $^{210}$Po are underestimated with respect to the 
experimental ones, but they are more sensitive to possible collective core 
excitations.
In fact, all over the trans-lead region the large observed $B(E3)$ values 
reflect the collective nature of the $3^-$ state at 2.615 MeV in $^{208}$Pb 
\cite{bergstrom85}.

Two quadrupole moments in $^{210}$Po and one in $^{206}$Hg are experimentally 
known: they are $Q(8^+_1)$ and $Q(11^-_1)$ in $^{210}$Po, and the 
$Q(5^-_1)$ in $^{206}$Hg.
Our calculated values are $-58.8$, $-97$, and $53 ~{\mathrm e} \cdot 
{\mathrm fm}^2$ to be compared with the experimental values 
\cite{blomqvist91,becker82} $-55.2 \pm 2.0$, $-86 \pm 11$, and 
$74 \pm 15 ~{\mathrm e} \cdot {\mathrm fm}^2$, respectively.

In Table II we compare the experimental $g$-factors \cite{browne92,helmer90} 
with the calculated ones.

\begin{table}[H]
\caption[]{Calculated and experimental $g$-factors. 
The experimental data are from \cite{browne92,helmer90}.}
\begin{center}
\renewcommand{\arraystretch}{1.2}
\begin{tabular}{lcccccc}
\hline\noalign{\smallskip}
 Nucleus &~& $J^{\pi}$ &~& \multicolumn{3}{c} {$g$} \\
 ~~~&~&~~~&~& Calc. &~& Expt. \\
\hline
 $^{210}$Po &~& $6^+_1$   &~& 0.906  &~& $0.913 \pm 0.006$ \\
 ~~~~~~~~~  &~& $8^+_1$   &~& 0.907  &~& $0.919 \pm 0.005$ \\
 ~~~~~~~~~  &~& $11^-_1$  &~& 1.140  &~& $1.108 \pm 0.012$ \\
 $^{206}$Hg &~& $5^-_1$   &~& 1.17   &~& $1.09 \pm 0.01$ \\
\noalign{\smallskip}\hline
\end{tabular}
\renewcommand{\arraystretch}{1}
\label{table2}
\end{center}
\end{table}

\section{Summary}
In summary, we have presented here the results of a shell-model study of the
$N=126$ isotones $^{210}$Po and $^{206}$Hg, where use has been made of 
effective two-particle and two-hole interactions derived from the Bonn A $NN$ 
potential. 
The agreement between theory and experiment is very good for both nuclei.
The data available on $^{206}$Hg are, however, rather scanty.
More experimental information on this nucleus is most desirable to put to a 
test the predictive power of our calculations.
It should be emphasized that, together with those of Ref. \cite{coraggio98}, 
these are the first shell-model calculations in the lead region where the 
effective interaction is derived from a modern $NN$ potential by means of a 
$G$-matrix folded-diagram method. 

In a forthcoming paper we shall present the results of an extensive study of 
the $N=126$ isotones \cite{coraggio99}.
Here, we conclude that the present results, which are consistent with those 
obtained in our previous works \cite{andr96,andr97a,andr97b,coraggio98}, 
provide further insight into the role of modern realistic interactions in 
nuclear structure calculations, evidencing, in particular, the reliability of 
the Bonn potential.

\section*{Acknowledgments}
The results presented in this paper are part of a research project carried out 
in collaboration with T. T. S. Kuo.
This work was supported in part by the Italian Ministero dell'Universit\`a e 
della Ricerca Scientifica e Tecnologica (MURST). 



\begin{thebibliography}{99}
\bibitem{autoc} We define $\sigma=\{ (1/N_d) \sum_i [ E_{\mathrm{expt}}(i) - 
E_{\mathrm{calc}}]^2 \}^{1/2}$, where $N_d$ is the number of data.
\end{thebibliography}
\end{document}